\documentclass[10pt]{article}
\usepackage{graphicx}
\usepackage{amsmath}
\usepackage{amssymb}
\usepackage{caption2}
\begin{document}
\bibliographystyle{prsty}
\begin{center}
{\large {\bf \sc{  Analysis of the $\overline{\Xi}_{cc}\Xi_{cc}$ hexaquark molecular state with the QCD sum rules }}} \\[2mm]
Zhi-Gang  Wang \footnote{E-mail: zgwang@aliyun.com.  }   \\
 Department of Physics, North China Electric Power University, Baoding 071003, P. R. China
\end{center}

\begin{abstract}
In this work, we construct the color-singlet-color-singlet type six-quark pseudoscalar current to investigate
 the $\overline{\Xi}_{cc}\Xi_{cc}$ hexaquark molecular state with the QCD sum rules,
 the predicted mass $M_X \sim 7.2\,\rm{GeV}$ supports assigning the $X(7200)$ to be the $\overline{\Xi}_{cc}\Xi_{cc}$ hexaquark molecular state
 with the quantum numbers $J^{PC}=0^{-+}$. The $X(7200)$ can decay through fusions of the $\bar{c}c$
and $\bar{q}q$ pairs, we can search for the $X(7200)$ and explore its properties in the  $J/\psi J/\psi$, $\chi_{c1}\chi_{c1}$, $D^*\bar{D}^*$ and
$D_1\bar{D}_1$ invariant mass spectrum in the future.
\end{abstract}



\section{Introduction}
In 2017, the LHCb collaboration observed the doubly charmed baryon state  $\Xi_{cc}^{++}$ in the $\Lambda_c^+ K^- \pi^+\pi^+$ invariant mass distribution \cite{LHCb-Xicc}.
The observation of the $\Xi_{cc}^{++}$ makes a great progress on the spectroscopy of the doubly charmed  baryon states, tetraquark states and pentaquark states.

 In 2020, the LHCb collaboration  investigated  the $J/\psi J/\psi$   invariant mass distributions  and observed a
 narrow structure about $6.9\, \rm{GeV}$  and   a broad structure just above the $J/\psi J/\psi$ threshold  at $p_T> 5.2\,\rm{GeV}$  with the global significances larger
 than $5\sigma$ \cite{LHCb-cccc-2006}. The Breit-Wigner mass and width of the $X(6900)$ are
$M_X=6905 \pm 11 \pm 7\, \rm{MeV}$ and $\Gamma_{X}=80 \pm 19 \pm  33\, \rm{MeV}$, respectively. Furthermore, they also observed  some vague structures around
$7.2\,\rm{ GeV}$, which coincides with the $\overline{\Xi}_{cc}\Xi_{cc}$ threshold $7242.4\,\rm{MeV}$ \cite{LHCb-cccc-2006}.
  The energy is sufficient to create a  baryon-antibaryon pair $\overline{\Xi}_{cc}\Xi_{cc}$ containing the valence quarks
$ccq \, \bar c \bar c \bar q$.

In the dynamical diquark model, the first radial excited states of the D-wave tetraquark states with the valence
quarks $cc\bar{c}\bar{c}$ have the masses about $7.2\,\rm{GeV}$,
 the insignificant enhancement $X(7200)$ may be a combination of some 2P and (or) 2D diquark-antidiquark type $cc \bar{c}\bar{c}$ states  with threshold effects
  of the transitions  $\Xi_{cc}\overline{\Xi}_{cc} \to J/\psi J/\psi$ \cite{Lebed-XiXi-7200}. The assignment of the $Y(4630)$ serves as a benchmark, in this model,
the $Y(4630)$ can be assigned to be a diquark-antidiquark type tetraquark state, fragmentation of the color
flux-tube connecting the diquark-antidiquark pair $cq\bar{c}\bar{q}$ leads to the lowest-lying baryon-antibaryon pair $\Lambda_c^+\Lambda_c^-$ \cite{Brodsky-Lebed-PRL}.
 Just like the $Y(4630)$, fragmentation of the color
flux-tube connecting the  diquark-antidiquark pair $cc\bar{c}\bar{c}$ in the $X(7200)$ leads to the lowest-lying baryon-antibaryon pair $\Xi_{cc}\overline{\Xi}_{cc}$,
then translates to the $J/\psi J/\psi$ pair.

 In the V-baryonium tetraquark scenario and the string-junction scenario, which share the same feature, the exotic $X$, $Y$ and $Z$ states are
 genuine tetraquarks rather than  molecular states,  they have  a baryonic vertex (or string-junction) attaching a $cc$-diquark in color antitriplet,
  which is connected by a string to
 an anti-baryonic vertex (or string-junction) attaching
 a $\bar{c}\bar{c}$-antidiquark in color triplet. In the V-baryonium tetraquark scenario, the un-conformed structure  $X(7200)$ is assigned to be  the first
 radially excited state of the $X(6900)$, the decays
  to the baryon-antibaryon pair $\Xi_{cc}\overline{\Xi}_{cc}$ can occur via breaking the string and creating a quark-antiquark pair
   \cite{Weissman-7200}. In the string junction scenario, the $X(6900)$ and $X(7200)$ are assigned to be the 2S tetraquark states with the quantum numbers $J^{PC}=0^{++}$ and $2^{++}$,
   respectively \cite{Karliner-Rosner-7200}.

If the $\overline{\Xi}_{cc}^{(\ast)}\Xi_{cc}^{(\ast)}$ system and $D^{(\ast)}\bar{D}^{(\ast)}$ system  are related to each other  via heavy antiquark-diquark symmetry,
 several molecular states can be predicted  based on the contact-range effective field theory, the $\overline{\Xi}_{cc}\Xi_{cc}$ molecular states which have the quantum numbers
  $J^{PC}=0^{-+}$ and (or) $1^{--}$ maybe contribute to the $X(7200)$ \cite{LSGeng-7200}.

The exotic $X$, $Y$, $Z$ and $P$ states always lie near two-particle thresholds, such as
\begin{eqnarray}
D\bar{D}^*/\bar{D}D^* &:& X(3872)\, ,\, Z_c(3885/3900)\, , \nonumber \\
D^*\bar{D}^*&:& Z_c(4020/4025) \, , \nonumber\\
 D\bar{D}_s^*/D^*\bar{D}_s &:& Z_{cs}(3985/4000) \, , \nonumber\\
 D_s^*\bar{D}_s^*&:& X(4140) \, , \nonumber\\
 D\bar{D}_1/\bar{D}D_1 &:& Y(4260/4220)\, , \, Z_c(4250)\, , \nonumber \\
D^*\bar{D}_0/\bar{D}^*D_0 &:& Y(4360/4320)\, , \nonumber \\
  \bar{D}\Sigma_c&:& P_c(4312) \, ,\nonumber \\
 \bar{D}\Xi^\prime_c/\bar{D}^*\Xi_c &:& P_{cs}(4459) \, ,\nonumber \\
  \bar{D}\Sigma^*_c&:& P_c(4380) \, ,\nonumber \\
  \bar{D}^*\Sigma_c&:& P_c(4440/4457) \, ,\nonumber \\
  \Lambda_c^+\Lambda_c^{-}/f_0(980)\psi^\prime &:& Y(4630/4660)\, ,\nonumber\\
B\bar{B}^*/\bar{B}B^* &:& Z_b(10610)\, , \nonumber \\
B^*\bar{B}^*&:& Z_b(10650) \, ,
\end{eqnarray}
it is natural to assume that they are molecular states composed of two color-singlet constituents and investigate their properties with the QCD sum rules
\cite{Narison-IJMPA-mole,Nielsen-DsD-mole,WZG-penta-mole,KAzizi-mole,WZG-mole-EPJC-formula,ChenHX-mole}.
The QCD sum rules approach is powerful theoretical tool in exploring the hadron properties, and has given many successful descriptions of the masses and decay widths of the
tetraquark and pentaquark (molecular) states \cite{Narison-IJMPA-mole,Nielsen-DsD-mole,WZG-penta-mole,KAzizi-mole,WZG-mole-EPJC-formula,ChenHX-mole},
and has been successfully applied  to investigate the dibaryon states and baryonium states \cite{WZG-triply-charm-dibaryon,QiaoCF-mole-dibaryon}.

In the present work, we explore the color-singlet-color-singlet type hexaquark molecular state $\overline{\Xi}_{cc}\Xi_{cc}$ with the quantum numbers $J^{PC}=0^{-+}$,
and make possible assignment of the $X(7200)$.

The article is arranged as follows:  we derive the QCD sum rules for the mass and pole residue
of the  $\overline{\Xi}_{cc}\Xi_{cc}$ molecular state in section 2; in section 3, we present the numerical results and discussions; section 4 is reserved for our conclusion.

\section{QCD sum rules for  the  $\overline{\Xi}_{cc}\Xi_{cc}$  hexaquark molecular  state  }
Let us  write down  the two-point correlation function   $\Pi (p^2)$ firstly,
\begin{eqnarray}
\Pi(p^2)&=&i\int d^4x e^{ip \cdot x} \langle0|T\left\{J(x)J^{\dagger}(0)\right\}|0\rangle \, ,
\end{eqnarray}
where
\begin{eqnarray}
J(x)&=&\bar{J}_{cc}(x) i\gamma_5 J_{cc}(x)\, , \nonumber\\
J_{cc}(x)&=&\varepsilon^{ijk} c^T_{i}(x) C\gamma_\alpha c_j(x)\gamma^\alpha\gamma^5 q(x)\, ,
\end{eqnarray}
$q=u$, $d$,  the $i$, $j$, $k$ are color indexes, the current $J_{cc}(x)$ has the same quantum numbers as the doubly charmed baryon state  $\Xi_{cc}$, and couples potentially to
the color-singlet clusters  with the same quantum numbers as the $\Xi_{cc}$ \cite{WZG-Xicc}.

At the hadron  side, we isolate the ground state contribution of the pseudoscalar hexaquark  molecular  state, we obtain the result,
\begin{eqnarray}\label{CF-hadron-side}
\Pi (p^2) &=&\frac{\lambda_X^2}{M^2_X-p^2} +\cdots \, \, ,
\end{eqnarray}
where the pole residue  $\lambda_{X}$ is defined by
\begin{eqnarray}
 \langle 0|J (0)|X (p)\rangle &=& \lambda_{X}     \, .
\end{eqnarray}
In Eq.\eqref{CF-hadron-side}, we have neglected the two-particle scattering state contributions, just like in the QCD sum rules for the tetraquark (molecular) states.
In Ref.\cite{ChuSheng-3-Nc},
 Lucha,  Melikhov and Sazdjian  obtain the conclusion that  "A possible exotic tetraquark state may appear only in $N_c$-subleading
 contributions to the QCD Green functions" based on the naive large-$N_c$ counting.
 An explicit analysis by S. Narison et al at the real word $N_c = 3$  shows the opposite
 due to the induced loop factor missed in a naive $N_c$ counting rule \cite{Narison-to-Chusheng-3}. In fact, as pointed in Ref.\cite{WZG-2102}, without excluding the contributions of
 factorizable Feynman diagrams in the color space to the QCD sum rules by hand, we cannot obtain the conclusion that the factorizable parts of
the operator product expansion series cannot have any relationship to the possible tetraquark
bound states.  For the baryon, tetraquark, pentaquark states with
string junctions \cite{Veneziano-Nc}, the  standard large-$N_c$ counting rules for the ordinary mesons cannot be trivially extrapolated
to the exotic hadrons but should be modified for being properly applied at the real word $N_c = 3$. All in all,
we can examine  the predictions of  the multiquark states  based on the QCD sum rules in different channels in the future.

In the QCD side, we carry out the operator product expansion up to the vacuum condensates of dimension 10 in a consistent way. There are
four heavy quark propagators and two light quark propagators in the correlation function $\Pi(p^2)$ after accomplishing the Wick's contractions.
If each heavy quark line emits a gluon, and each light quark line contributes a quark-antiquark pair,
we obtain a quark-gluon operator $g_sG_{\mu\nu}g_sG_{\alpha\beta}g_sG_{\rho\sigma}g_sG_{\lambda\tau}\bar{q}q\bar{q}q$, which is of dimension $14$, we should take account of the
vacuum condensates of dimensions up to dimension 14.
In the QCD sum rules for the hidden-charm or hidden-bottom tetraquark (molecular) states, pentaquark (molecular) states, we usually take the truncation
$\mathcal{O}(\alpha_s^k)$ with $k\leq 1$  \cite{WZG-penta-mole,WZG-mole-EPJC-formula,WangHuangtao-2014-PRD,WZG-IJMPA-penta}. If we also take the
truncation $k\leq 1$  to discard the quark-gluon operators of the orders
$\mathcal{O}(\alpha_s^{>1})$  in the present work,   the operator product expansion is terminated at the vacuum condensates of dimension $10$.

In the QCD sum rules for the  triply-charmed diquark-diquark-diquark type hexaquark states and triply-charmed dibaryon states,
there are three light quark propagators and three heavy quark propagators in the correlation functions after accomplishing the Wick's contractions
\cite{WZG-triply-charm-dibaryon,WZG-triply-charm-hexa}.
Again, if each heavy quark line emits a gluon and each light quark line contributes quark-antiquark pair, we obtain a quark-gluon
operator $g_sG_{\mu\nu}g_sG_{\alpha\beta}g_sG_{\lambda\tau}\bar{q}q\bar{q}q\bar{q}q$, which is of dimension 15.  If we take the truncations $k\leq 1$,
   the operator product expansion is terminated at the vacuum condensates of dimension $13$.
The operator $g_sG_{\mu\nu}g_sG_{\alpha\beta}g_sG_{\lambda\tau}\bar{q}q\bar{q}q\bar{q}q$ leads to the vacuum condensates
$\langle\frac{\alpha_sGG}{\pi}\rangle\langle\bar{q}q\rangle^2\langle\bar{q}g_s\sigma Gq\rangle$, $\langle g_s^3 GGG\rangle\langle\bar{q}q\rangle^3$
and $\langle\bar{q}g_s\sigma Gq\rangle^3$, we calculate the vacuum condensate $\langle\bar{q}g_s\sigma Gq\rangle^3$, and observe that its small contribution
 can be
neglected safely \cite{WZG-triply-charm-dibaryon,WZG-triply-charm-hexa}. The vacuum condensates
$\langle g_s^3 GGG\rangle\langle\bar{q}q\rangle^3$ and $\langle\frac{\alpha_sGG}{\pi}\rangle\langle\bar{q}q\rangle^2\langle\bar{q}g_s\sigma Gq\rangle$ receive additional
suppressions due to the small contributions of the gluon condensate and  three-gluon condensate. In Ref.\cite{WZG-AAPPS}, we re-explore
the mass spectrum of the ground state triply-heavy baryon states with the QCD sum rules by
 taking  account of the three-gluon condensate $\langle g_s^3 GGG\rangle$ for the first time, and observe that the contributions of the three-gluon condensate $\langle g_s^3 GGG\rangle$ are tiny indeed.
  In summary, the truncations $k\leq 1$ work very good.

In the present wok, we take  account of the vacuum condensates   $\langle\bar{q}q\rangle$,   $\langle\bar{q}g_{s}\sigma Gq\rangle$, $\langle\bar{q}q\rangle^2 $,
$\langle\bar{q}q\rangle  \langle\bar{q}g_{s}\sigma Gq\rangle$,
$\langle\frac{\alpha_{s}GG}{\pi}\rangle$,
$\langle\bar{q}q\rangle \langle\frac{\alpha_{s}GG}{\pi}\rangle$,
$\langle\bar{q}q\rangle^2 \langle\frac{\alpha_{s}GG}{\pi}\rangle$ and $\langle\bar{q}g_{s}\sigma Gq\rangle^2$,
which are the vacuum expectation values of the quark-gluon operators of the orders $\mathcal{O}(\alpha_s^k)$ with  $k\leq1$.
The vacuum condensates $\langle g_s^3GGG\rangle$, $\langle\frac{\alpha_sGG}{\pi}\rangle^2$,
$\langle\frac{\alpha_sGG}{\pi}\rangle \langle \bar{q}g_s\sigma G q\rangle$ and $\langle \bar{q}q\rangle\langle g_s^3GGG\rangle$ have the dimensions
 $6$, $8$, $9$ and $9$, respectively, however, they are the vacuum expectation values of the quark-gluon operators of the orders
 $\mathcal{O}(\alpha_s^{\frac{3}{2}})$, $\mathcal{O}(\alpha_s^{2})$, $\mathcal{O}(\alpha_s^{\frac{3}{2}})$  and $\mathcal{O}(\alpha_s^{\frac{3}{2}})$, respectively,
  and are  neglected, direct calculations indicate that those contributions are tiny indeed \cite{WangXW-Wang}.

 We obtain the spectral density at the quark level through dispersion relation, take the quark-hadron duality below the continuum threshold  $s_0$,
  and perform Borel transform  in regard to
the variable $P^2=-p^2$ to obtain  the QCD sum rules:
\begin{eqnarray}\label{QCDSR}
\lambda^2_{X}\, \exp\left(-\frac{M^2_{X}}{T^2}\right)&=& \int_{16m_c^2}^{s_0} ds  \, \rho_{QCD}(s)  \exp\left(-\frac{s}{T^2}\right) \, ,
\end{eqnarray}
where the $\rho_{QCD}(s)$ is the  spectral density at the quark level.

We derive   Eq.\eqref{QCDSR} with respect to  $\tau=\frac{1}{T^2}$, then eliminate the
 pole residue $\lambda_{X}$, and  obtain the QCD sum rules for
 the mass of the pseudoscalar hexaquark molecular  state $\overline{\Xi}_{cc}\Xi_{cc}$,
 \begin{eqnarray}\label{QCDSR-mass}
 M^2_{X}&=&- \frac{\frac{d}{d \tau}\int_{16m_c^2}^{s_0} ds  \, \rho_{QCD}(s)  \exp\left(-\frac{s}{T^2}\right)}{\int_{16m_c^2}^{s_0} ds  \, \rho_{QCD}(s)
  \exp\left(-\frac{s}{T^2}\right)}\, .
  \end{eqnarray}

\section{Numerical results and discussions}

We adopt  the standard values  of  the  vacuum condensates
$\langle\bar{q}q \rangle=-(0.24\pm 0.01\, \rm{GeV})^3$,  $\langle\bar{q}g_s\sigma G q \rangle=m_0^2\langle \bar{q}q \rangle$,
 $m_0^2=(0.8 \pm 0.1)\,\rm{GeV}^2$, $\langle \frac{\alpha_s
GG}{\pi}\rangle=0.012\pm0.004\,\rm{GeV}^4$    at the  energy scale  $\mu=1\, \rm{GeV}$
\cite{SVZ79,PRT85,Ioffe-NPB-1981,Ioffe-mixcondensate,ColangeloReview}, and  take the $\overline{MS}$ mass of the charm  quark, $m_{c}(m_c)=(1.275\pm0.025)\,\rm{GeV}$,
 from the Particle Data Group \cite{PDG}.
In addition,  we take  account of the energy-scale dependence of  all the input parameters \cite{Narison-mix},
 \begin{eqnarray}
 \langle\bar{q}q \rangle(\mu)&=&\langle\bar{q}q\rangle({\rm 1 GeV})\left[\frac{\alpha_{s}({\rm 1 GeV})}{\alpha_{s}(\mu)}\right]^{\frac{12}{33-2n_f}}\, , \nonumber\\
  \langle\bar{q}g_s \sigma Gq \rangle(\mu)&=&\langle\bar{q}g_s \sigma Gq \rangle({\rm 1 GeV})\left[\frac{\alpha_{s}({\rm 1 GeV})}{\alpha_{s}(\mu)}\right]^{\frac{2}{33-2n_f}}\, ,\nonumber\\
 m_c(\mu)&=&m_c(m_c)\left[\frac{\alpha_{s}(\mu)}{\alpha_{s}(m_c)}\right]^{\frac{12}{33-2n_f}} \, ,\nonumber\\
\alpha_s(\mu)&=&\frac{1}{b_0t}\left[1-\frac{b_1}{b_0^2}\frac{\log t}{t} +\frac{b_1^2(\log^2{t}-\log{t}-1)+b_0b_2}{b_0^4t^2}\right]\, ,
\end{eqnarray}
  where $t=\log \frac{\mu^2}{\Lambda^2}$, $b_0=\frac{33-2n_f}{12\pi}$, $b_1=\frac{153-19n_f}{24\pi^2}$, $b_2=\frac{2857-\frac{5033}{9}n_f+\frac{325}{27}n_f^2}{128\pi^3}$,  $\Lambda=213\,\rm{MeV}$, $296\,\rm{MeV}$  and  $339\,\rm{MeV}$ for the quark flavor numbers  $n_f=5$, $4$ and $3$, respectively  \cite{PDG}.
In the present work, we explore  the hidden-charm hexaquark molecular  state $\overline{\Xi}_{cc}\Xi_{cc}$ and choose $n_f=4$, then  evolve all the
input parameters  to a typical energy scale $\mu$ to extract the hexaquark molecule  mass.
Furthermore, we present the predictions  based on the updated parameters obtained by S. Narison,  $m_{c}(m_c)=(1.266\pm0.006)\,\rm{GeV}$ and $\langle \frac{\alpha_s
GG}{\pi}\rangle=0.021\pm0.001\,\rm{GeV}^4$ \cite{Narison-2101}, and in this case the energy scales of other vacuum condensates are taken at $\mu=1\,\rm{GeV}$.

\begin{table}
\begin{center}
\begin{tabular}{|c|c|c|c|c|c|c|c|}\hline\hline
   $J^{PC}$                    & 1S                 & 2S                      & energy gaps    & References      \\ \hline
   $1^{+-}$                    & $Z_c(3900)$        & $Z_c(4430)$             & 591\,\rm{MeV}  & \cite{Maiani-Z4430-1405,Nielsen-1401,WangZG-Z4430-CTP}      \\ \hline
    $0^{++}$                   & $X(3915)$          & $X(4500)$               & 588\,\rm{MeV}  & \cite{X4140-tetraquark-Lebed,X3915-X4500-EPJC-WZG,X3915-X4500-EPJA-WZG}  \\ \hline
    $1^{+-}$                   & $Z_c(4020)$        & $Z_c(4600)$             & 576\,\rm{MeV}  & \cite{ChenHX-Z4600-A,WangZG-axial-Z4600}  \\ \hline
   $1^{++}$                    & $X(4140)$          & $X(4685)$               & 566\,\rm{MeV}  & \cite{WZG-Y4140-Y4685}    \\ \hline \hline
\end{tabular}
\end{center}
\caption{ The energy gaps between the ground states (1S) and first radial excited states (2S) of
the hidden-charm tetraquark candidates with the possible assignments. }\label{1S2S}
\end{table}

We should choose suitable continuum threshold $s_0$ to avoid contamination from the first radial excited state.  In the scenario of the tetraquark states, the possible assignments of the exotic states $Z_c(3900)$, $Z_c(4430)$, $X(3915)$, $X(4500)$,
$Z_c(4020)$,  $Z_c(4600)$, $X(4140)$ and  $X(4685)$ are presented plainly  in Table \ref{1S2S} according to the (possible) quantum numbers, decay modes and energy gaps. From the table,
we can obtain the conclusion tentatively that the energy gaps between the ground states and first radial excited states
of the hidden-charm tetraquark states are about $0.58\,\rm{GeV}$. We can  choose the continuum threshold parameter as $\sqrt{s_0}= M_{X}+0.6\pm0.1\,\rm{GeV}$.
If the mass of the $\overline{\Xi}_{cc}\Xi_{cc}$ hexaquark molecular state and the energy scale of the spectral density at the quark level satisfies energy scale formula
$M_X=\sqrt{\mu^2+(4\mathbb{M}_c)^2}$, the lower bound of the mass $M_X\geq\sqrt{(1\,\rm{GeV})^2+(4\times 1.85\,\rm{GeV})^2}=7.47\,\rm{GeV}> 2M_{\Xi_{cc}}$.

Now let us suppose that the multiquark states $X$, $Y$, $Z$ and $P$ have $N_Q+N_q$ valence quarks, where the $N_Q$ and $N_q$ are the
 numbers of the heavy quarks and light quarks, respectively.
Generally speaking, if $N_Q\leq N_q$, we can apply the energy scale formula $\mu=\sqrt{M_{X/Y/Z/P}^2-(N_Q \mathbb{M}_Q)^2}$ to enhance the pole contributions and improve the
convergent behavior of the operator product expansion
\cite{WZG-penta-mole,WZG-mole-EPJC-formula,WZG-triply-charm-dibaryon,WangHuangtao-2014-PRD,WZG-IJMPA-penta,WZG-triply-charm-hexa}.
In the present case, $N_Q=4> N_q=2$, the energy scale formula is not applicable.

 After  trial and error,   we  obtain the continuum threshold parameter $\sqrt{s_0}=7.8\pm 0.1\,\rm{GeV}$, and  Borel parameters which are shown
 in Table \ref{Borel} for four typical energy scales $\mu=1.0\,\rm{GeV}$, $m_c(m_c)$, $1.5\,\rm{GeV}$, $2.0\,\rm{GeV}$ and $1.0 \,\rm{GeV}^*$,
 where the superscript $*$ denotes the $c$-quark mass $m_{c}(m_c)=(1.266\pm0.006)\,\rm{GeV}$ and gluon condensate $\langle \frac{\alpha_s
GG}{\pi}\rangle=0.021\pm0.001\,\rm{GeV}^4$ are taken for Ref.\cite{Narison-2101}. In the Borel windows,
 the ground state contributions are about $(18-37)\%$ or $(22-41)\%$ and the pole contribution cannot reach $50\%$,
   the contributions of the vacuum condensates of dimension 10 are about $(2-6)\%$,  $< 1\%$, $< 1\%$, $<1\%$ and $(1-2)\%$, respectively,
 the operator product expansion  converges  very well.

\begin{figure}
 \centering
 \includegraphics[totalheight=5cm,width=7cm]{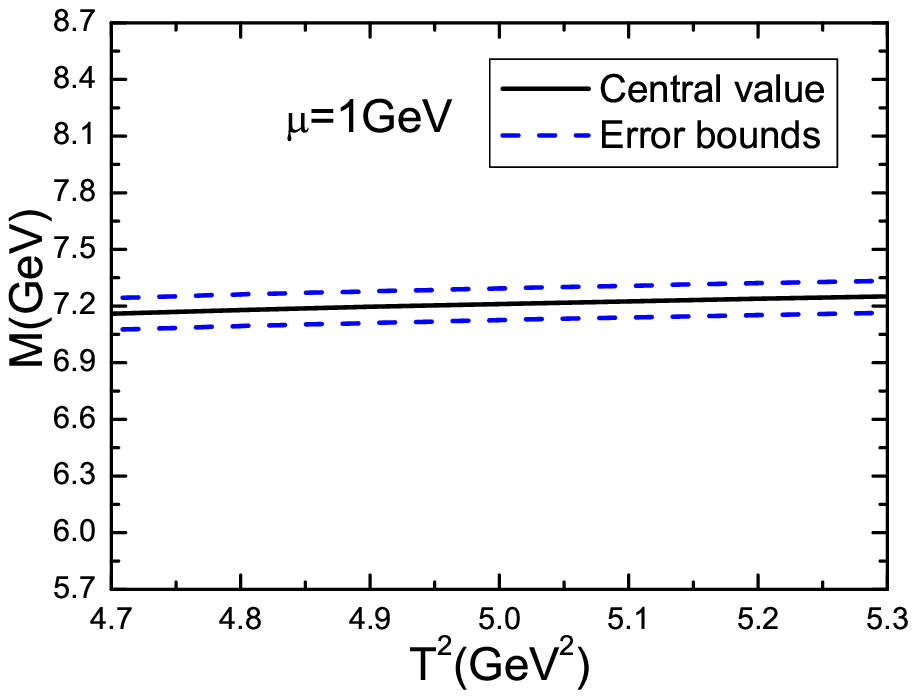}
 \includegraphics[totalheight=5cm,width=7cm]{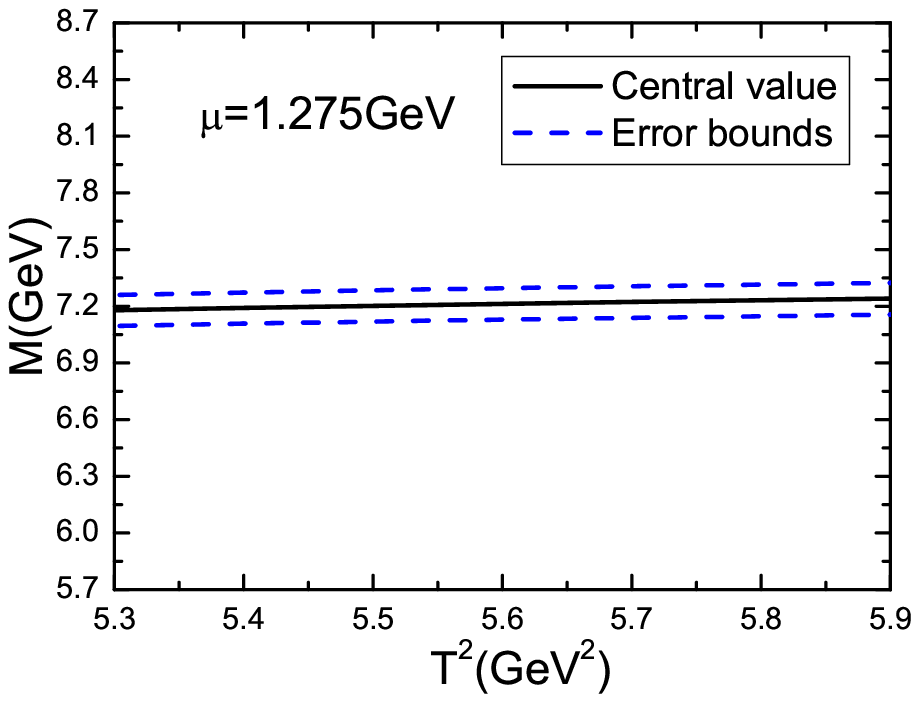}
 \includegraphics[totalheight=5cm,width=7cm]{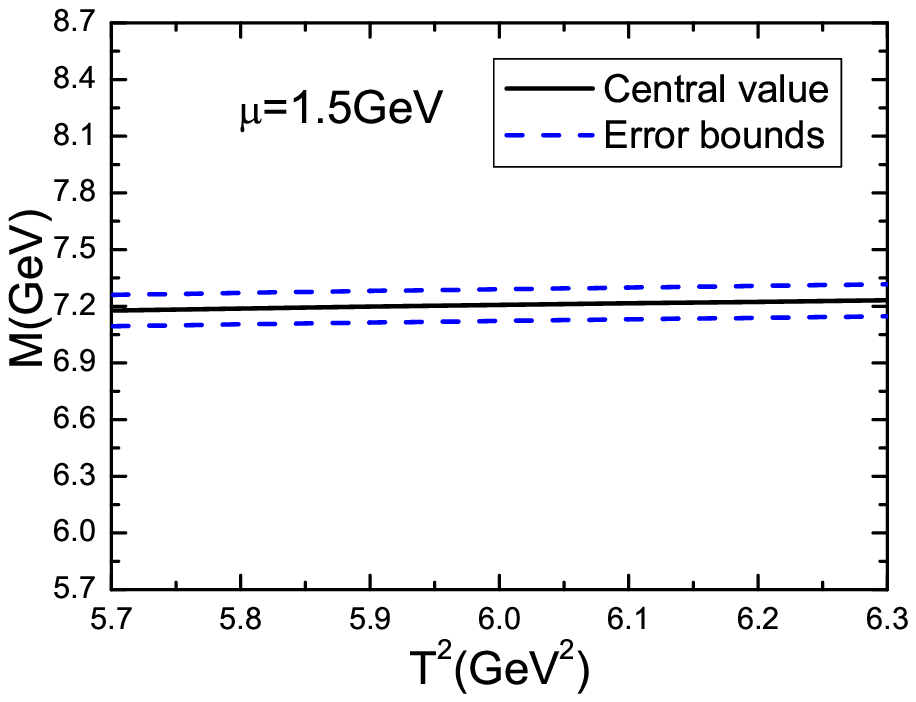}
 \includegraphics[totalheight=5cm,width=7cm]{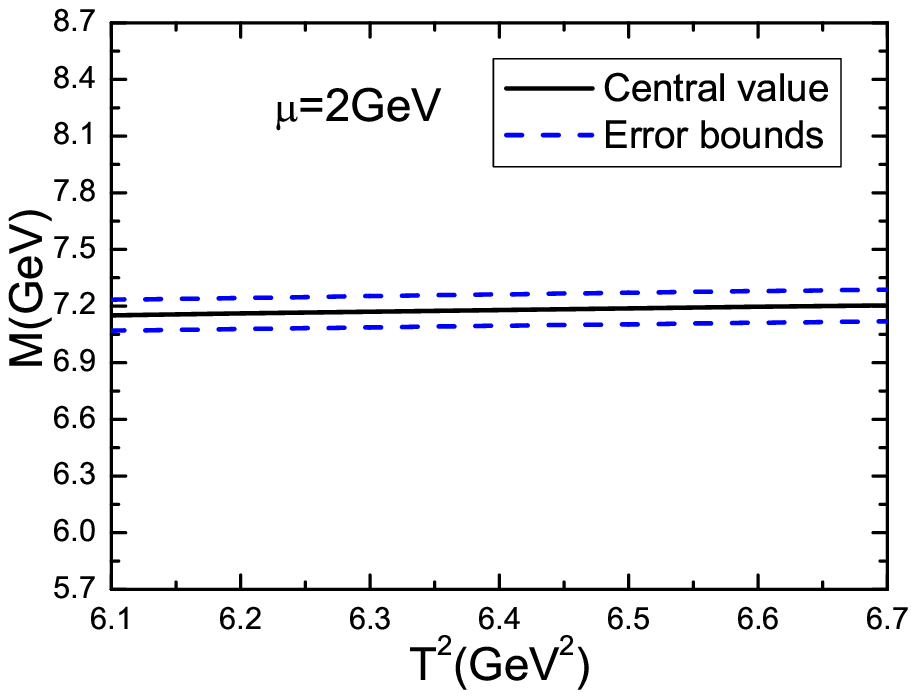}
 \includegraphics[totalheight=5cm,width=7cm]{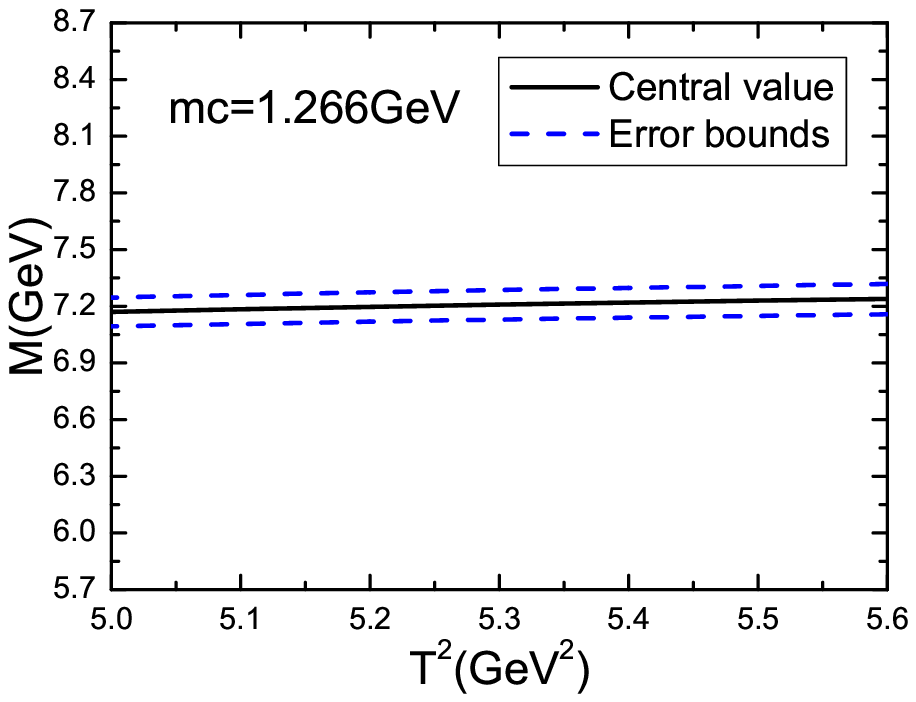}
  \caption{ The mass   of the $\overline{\Xi}_{cc}\Xi_{cc}$  hexaquark molecular state  with variation of the Borel parameter $T^2$,
  where the  $m_{c}(m_c)=1.266\,\rm{GeV}$ denotes the $c$-quark mass and gluon condensate  are taken from Ref.\cite{Narison-2101}.   }\label{mass-cccc}
\end{figure}

Then we  take account of all uncertainties of the  parameters,  and obtain the values of the mass and pole residue of the
pseudoscalar hexaquark molecular state $\overline{\Xi}_{cc}\Xi_{cc}$, which are  shown explicitly in Table \ref{Borel} and Fig.\ref{mass-cccc}.
 From Fig.\ref{mass-cccc}, we can see that the predicted mass   is rather stable with variation of the Borel parameter, the uncertainty comes
  from the Borel parameter  is rather  small. From Table \ref{Borel}, we can see that the predicted mass $M_X$ is almost independent on the
  energy scale of the QCD spectral density,
  while the pole residue depends heavily on the energy scale of the QCD spectral density, which
  is qualitatively consistent with the evolution behavior  of the current operator from the re-normalization group equation,
  $J(x,\mu)=L^{\gamma_{J}} J(x,\mu_0)$ and $\lambda_X(\mu)=L^{\gamma_{J}} \lambda_X(\mu_0)$,
  where $L=\frac{\alpha_s(\mu_0)}{\alpha_s(\mu)}$, and the $\gamma_J$ is the anomalous dimension of the current operator  $J(x)$. The predicted mass $M_X \sim 7.2\,\rm{GeV}$ is compatible with
  the  vague structure around $7.2\,\rm{GeV}$ in the $J/\psi J/\psi$ invariant  mass   spectrum observed by the LHCb collaboration \cite{LHCb-cccc-2006}, and supports assigning the $X(7200)$ as
  the $\overline{\Xi}_{cc}\Xi_{cc}$ hexaquark molecular state with the quantum numbers $J^{PC}=0^{-+}$. On the other hand, direct calculations based on the QCD sum rules \cite{WZG-QQQQ}
  do not support assigning the $X(7200)$ as the first radial excited state of the diquark-antidiquark-type (${\bf \bar{3}}_c{\bf 3}_c $- type) $cc\bar{c}\bar{c}$ tetraquark state
  claimed in Refs.\cite{Lebed-XiXi-7200,Weissman-7200,Karliner-Rosner-7200}.

\begin{table}
\begin{center}
\begin{tabular}{|c|c|c|c|c|c|c|c|}\hline\hline
$\mu(\rm GeV)$        &$T^2(\rm{GeV}^2)$      &pole          &$M_{X/Y}(\rm{GeV})$   &$\lambda(10^{-3}\rm{GeV}^8)$ \\ \hline

$1$                   &$4.7-5.3$              &$(18-37)\%$   &$7.21\pm0.13$         &$6.35 \pm 2.19$  \\ \hline

$1.275$               &$5.3-5.9$              &$(18-35)\%$   &$7.21\pm0.11$         &$13.1\pm 3.5$  \\ \hline

$1.5$                 &$5.7-6.3$              &$(17-32)\%$   &$7.21\pm0.12$         &$18.4\pm 4.6$  \\ \hline

$2$                   &$6.1-6.7$              &$(18-32)\%$   &$7.18\pm0.11$         &$27.9\pm 6.2$  \\ \hline

$1^*$                 &$5.0-5.6$              &$(22-41)\%$   &$7.21\pm0.12$         &$13.3\pm 2.8$  \\ \hline\hline

\end{tabular}
\end{center}
\caption{ The energy scale of the QCD spectral density,  Borel parameter, pole contribution, mass and pole residue of the hexaquark molecular
state $\overline{\Xi}_{cc}\Xi_{cc}$, where the superscript $*$ denotes the $c$-quark mass  and gluon condensate  are taken from  Ref.\cite{Narison-2101}. }\label{Borel}
\end{table}

The decays of the $\overline{\Xi}_{cc}\Xi_{cc}$ hexaquark molecule candidate $X(7200)$ can take place through fusions of the $\bar{c}c$
and $\bar{q}q$ pairs,
\begin{eqnarray}
X(7200)&\to &\overline{\Xi}_{cc}\Xi_{cc} \to  J/\psi J/\psi\, , \, \chi_{c1}\chi_{c1}\, ,\, D^*\bar{D}^*\, ,\, D_1\bar{D}_1\, ,
\end{eqnarray}
we can search for the $\overline{\Xi}_{cc}\Xi_{cc}$ hexaquark molecular state in the  $J/\psi J/\psi$, $\chi_{c1}\chi_{c1}$, $D^*\bar{D}^*$ and
$D_1\bar{D}_1$ invariant mass spectrum  at the BESIII, LHCb, Belle II,  CEPC, FCC and ILC in the future.

\section{Conclusion}
In the present work, we construct the color-singlet-color-singlet type six-quark pseudoscalar current to interpolate the $\overline{\Xi}_{cc}\Xi_{cc}$ hexaquark molecular state, then accomplish
the operator product expansion by calculating the vacuum condensates up to dimension 10, which are vacuum expectation values of the quark-gluon operators of the orders $\mathcal{O}(\alpha_s^k)$
with $k\leq 1$, and obtain the  QCD sum rules for the mass and pole residue. The predicted mass $M_X \sim 7.2\,\rm{GeV}$ supports assigning
the $X(7200)$ to be the $\overline{\Xi}_{cc}\Xi_{cc}$ hexaquark molecular state with the quantum numbers $J^{PC}=0^{-+}$. Moreover, direct calculations based on the QCD sum rules
  do not support assigning the $X(7200)$ as the first radial excited state of the diquark-antidiquark-type (${\bf \bar{3}}_c{\bf 3}_c $- type)
  $cc\bar{c}\bar{c}$ tetraquark state.
The decays of the $X(7200)$ can take place through fusions of the $\bar{c}c$
and $\bar{q}q$ pairs, we can search for the $\overline{\Xi}_{cc}\Xi_{cc}$ hexaquark molecular state and explore its properties in the  $J/\psi J/\psi$, $\chi_{c1}\chi_{c1}$, $D^*\bar{D}^*$ and
$D_1\bar{D}_1$ invariant mass spectrum  at the BESIII, LHCb, Belle II,  CEPC, FCC and ILC in the future.

\section*{Acknowledgements}
This  work is supported by National Natural Science Foundation, Grant Number  11775079.


\begin{thebibliography}{99}

\bibitem{LHCb-Xicc} R. Aaij  et al,  Phys. Rev. Lett. {\bf 119} (2017)  112001.


\bibitem{LHCb-cccc-2006} R. Aaij et al, Sci. Bull. {\bf 65} (2020)  1983.


\bibitem{Lebed-XiXi-7200} J. F. Giron and R. F. Lebed, Phys. Rev. {\bf D102} (2020)  074003.

\bibitem{Brodsky-Lebed-PRL} S. J. Brodsky, D. S. Hwang and R. F. Lebed, Phys. Rev. Lett. {\bf 113} (2014)  112001.




\bibitem{Weissman-7200} J. Sonnenschein and D. Weissman, Eur. Phys. J. {\bf C81} (2021)  25.


\bibitem{Karliner-Rosner-7200} M. Karliner and J. L. Rosner,  Phys. Rev. {\bf D102} (2020)  114039.

\bibitem{LSGeng-7200} M. Z. Liu and L. S. Geng, Eur. Phys. J. {\bf C81} (2021)  179.


\bibitem{Narison-IJMPA-mole} R. M. Albuquerque, F. Fanomezana, S. Narison and A. Rabemananjara, Phys. Lett. {\bf B715} (2012) 129;
R. Albuquerque, S. Narison, F. Fanomezana, A. Rabemananjara, D. Rabetiarivony and G. Randriamanatrika,
 Int. J. Mod. Phys. {\bf A31} (2016)  1650196.

\bibitem{Nielsen-DsD-mole} S. H. Lee, M. Nielsen and U. Wiedner, J. Korean Phys. Soc. {\bf 55} (2009) 424;
J. M. Dias, F. S. Navarra, M. Nielsen and C. M. Zanetti, Phys. Rev. {\bf D88} (2013) 016004.

\bibitem{WZG-penta-mole} Z. G. Wang, Int. J. Mod. Phys. {\bf A34} (2019)  1950097;
Z. G. Wang and X. Wang, Chin. Phys. {\bf C44} (2020) 103102.


\bibitem{JRZhang-mole} J. R. Zhang and M. Q. Huang, J. Phys. {\bf G37} (2010) 025005;
J. R. Zhang, M. Zhong and M. Q. Huang, Phys. Lett. {\bf B704} (2011) 312.


\bibitem{KAzizi-mole} K. Azizi, Y. Sarac and H. Sundu, Phys. Rev. {\bf D95} (2017)  094016;
K. Azizi, Y. Sarac and H. Sundu, Phys. Lett. {\bf B782} (2018) 694.

\bibitem{WZG-mole-EPJC-formula} Z. G. Wang and T. Huang, Eur. Phys. J. {\bf C74} (2014)  2891;
Z. G. Wang, Eur. Phys. J. {\bf C74} (2014)  2963.

\bibitem{ChenHX-mole} H. X. Chen, W. Chen, X. Liu, T. G. Steele and S. L. Zhu, Phys. Rev. Lett. {\bf 115} (2015) 172001;
W. Chen, T. G. Steele, H. X. Chen and S. L. Zhu, Phys. Rev. {\bf D92} (2015)  054002.



\bibitem{WZG-triply-charm-dibaryon} Z. G. Wang, Phys. Rev. {\bf D102} (2020)  034008.

\bibitem{QiaoCF-mole-dibaryon} B. D. Wan, L. Tang and C. F. Qiao, Eur. Phys. J. {\bf C80} (2020) 121.

\bibitem{WZG-Xicc} Z. G. Wang, Eur. Phys. J. {\bf C78} (2018) 826.


\bibitem{ChuSheng-3-Nc} W. Lucha, D. Melikhov and H. Sazdjian, Phys. Rev. {\bf D103} (2021)  014012.

\bibitem{Narison-to-Chusheng-3} R. M. Albuquerque, S. Narison and D. Rabetiarivony, Phys. Rev. {\bf D103} (2021)  074015.

\bibitem{WZG-2102} Z. G. Wang, arXiv:2102.07520 [hep-ph].

\bibitem{Veneziano-Nc} G. C. Rossi, T. Vergata and G. Veneziano, arXiv:2011.09774 [hep-ph].


\bibitem{WangHuangtao-2014-PRD} Z. G. Wang and T. Huang,  Phys. Rev. {\bf D89} (2014)  054019;
Z. G. Wang, Phys. Rev. {\bf D102} (2020)  014018.

\bibitem{WZG-IJMPA-penta} Z. G. Wang,  Int. J. Mod. Phys. {\bf A35} (2020)  2050003;
Z. G. Wang,  Int. J. Mod. Phys. {\bf A36} (2021) 2150071.


\bibitem{WZG-triply-charm-hexa} Z. G. Wang, Int. J. Mod. Phys. {\bf A35} (2020)  2050073.


\bibitem{WZG-AAPPS}   Z. G. Wang, AAPPS Bull. {\bf 31} (2021) 5.

\bibitem{WangXW-Wang} X. W. Wang and Z. G. Wang, in preparation.

\bibitem{SVZ79}  M. A. Shifman, A. I. Vainshtein and V. I. Zakharov, Nucl. Phys. {\bf B147} (1979) 385, 448.

\bibitem{PRT85} L. J. Reinders, H. Rubinstein and S. Yazaki, Phys. Rept. {\bf 127} (1985) 1.

\bibitem{Ioffe-NPB-1981} B. L. Ioffe, Nucl. Phys. {\bf B188} (1981) 317; Erratum: Nucl.Phys. {\bf B191} (1981) 591.

\bibitem{Ioffe-mixcondensate} V. M. Belyaev and B. L. Ioffe, Sov. Phys. JETP {\bf 56} (1982) 493.

\bibitem{ColangeloReview} P. Colangelo and A. Khodjamirian, hep-ph/0010175.

\bibitem{PDG} P. A. Zyla et al, Prog. Theor. Exp. Phys. {\bf 2020}  (2020) 083C01.


\bibitem{Narison-mix} S. Narison and R. Tarrach, Phys. Lett. {\bf 125 B} (1983) 217.

\bibitem{Narison-2101} S. Narison, arXiv:2101.12579.



\bibitem{Maiani-Z4430-1405} L. Maiani, F. Piccinini, A. D. Polosa and V. Riquer, Phys. Rev. {\bf D89} (2014) 114010.


\bibitem{Nielsen-1401} M. Nielsen and F. S. Navarra,  Mod. Phys. Lett. {\bf  A29} (2014) 1430005.

\bibitem{WangZG-Z4430-CTP} Z. G. Wang,  Commun. Theor. Phys. {\bf 63} (2015)  325.


\bibitem{X4140-tetraquark-Lebed} R. F. Lebed and A. D. Polosa,  Phys. Rev. {\bf D93} (2016) 094024.


\bibitem{X3915-X4500-EPJC-WZG} Z. G. Wang, Eur. Phys. J. {\bf C77} (2017)  78.

\bibitem{X3915-X4500-EPJA-WZG}   Z. G. Wang,  Eur. Phys. J. {\bf A53} (2017) 19.


\bibitem{ChenHX-Z4600-A} H. X. Chen and W. Chen,  Phys. Rev. {\bf D99} (2019)  074022.


\bibitem{WangZG-axial-Z4600} Z. G. Wang, Chin. Phys. {\bf C44} (2020) 063105.


\bibitem{WZG-Y4140-Y4685} Z. G. Wang, arXiv:2103.04236 [hep-ph].

\bibitem{WZG-QQQQ} Z. G. Wang and Z. Y. Di, Acta Phys. Polon. {\bf B50} (2019) 1335;
Z. G. Wang, Chin. Phys. {\bf C44} (2020) 113106;
Z. G. Wang, Int. J. Mod. Phys. {\bf A36} (2021)  2150014.

\end{thebibliography}
\end{document}